\documentstyle[aps,preprint]{revtex}
\begin{document}
\draft
\title{Effective Field Theory of ideal-fluid Hydrodynamics}
\author{Adriaan M. J. Schakel}
\address{Institut f\"ur Theoretische Physik \\ Freie Universit\"at Berlin \\
Arnimallee 14, 14195 Berlin}
\date{July 14, 1996}
\maketitle
\begin{abstract}
Starting from a standard description of an ideal, isentropic fluid, we
derive the effective theory governing a gapless non-relativistic
mode---the sound mode.  The theory, which is dictated by
the requirement of Galilei invariance, entails the entire set of
hydrodynamic equations.  The gaplessness of the sound mode is
explained by identifying it as the Goldstone mode associated with the
spontaneous breakdown of Galilei invariance.  Differences with a
superfluid are pointed out.
\end{abstract}
\pacs{}
\noindent
The hydrodynamics of an ideal, classical fluid was already well understood in
the 19th century.  The case of isentropic flow, for which the entropy per unit
mass is constant, is particularly simple.  The pressure $P$ is here a function
of the mass density $\rho$ only, and the flow is automatically a potential
flow.  A property of such a fluid is that it supports unattenuated sound
waves, i.e., propagating density oscillations.  The waves are unattenuated
because viscosity and thermal conductivity, which usually serve to dissipate
the energy of a propagating mode, are absent.  Sound waves belong to the class
of modes having a gapless energy spectrum.  The purpose of this essay is to
argue that their presence in an isentropic fluid is an {\em emergent} property
\cite{PWA}.  That is, we do not take the existence of this gapless mode for
granted, but wish to explain it from an underlying principle, namely that of
broken symmetry.

To describe the hydrodynamics of an isentropic  fluid, we use
Eckart's variational principle \cite{Eckart} and start with the
Lagrange density
\begin{equation} 
\label{L}
{\cal L} = \case{1}{2} \rho {\bf v}^2 - \rho e + \theta [\partial_0
\rho + \nabla \cdot (\rho {\bf v})],
\end{equation}
where ${\bf v}$ is the velocity field, $\rho$ the mass density, and $e$ the
internal energy per unit mass.  For isentropic flow $e$ is a function of
$\rho$ alone.  The first and second term in (\ref{L}) represent the kinetic
and potential energy density, respectively.  The variable $\theta$ is a
Lagrange multiplier introduced to impose the conservation of mass:
\begin{equation}  \label{mass}
\partial_0 \rho + \nabla \cdot (\rho {\bf v}) = 0;
\end{equation} 
its dimension is $[\theta] = {\rm m}^2 {\rm s}^{-1}$.  
The variation of (\ref{L}) with respect to ${\bf v}$ yields the equation
\begin{equation} 
\label{v}
{\bf v} = \nabla \theta.
\end{equation}
It shows that, indeed, isentropic flow is automatically a potential flow and
it also identifies the Lagrange multiplier $\theta$ as the velocity potential.
(How to incorporate vortices in this variational approach will be discussed
below.)  With (\ref{v}), the Lagrange density (\ref{L}) becomes
\begin{equation} 
\label{Ltheta}
{\cal L} = - \rho [\partial_0 \theta + \case{1}{2} (\nabla \theta)^2 +
e],
\end{equation}
where we performed integrations by part.  A second field equation can be
obtained by varying the Lagrange density with respect to $\rho$.  This yields
the Bernoulli equation
\begin{equation} 
\label{rhovar}
\partial_0 \theta + \case{1}{2} (\nabla \theta)^2 + h  = 0,
\end{equation} 
with $h = \partial (\rho e)/ \partial \rho$ the specific enthalpy.  On taking
the gradient of (\ref{rhovar}), and using that for isentropic flow one has the 
thermodynamic relation 
\begin{equation} 
\nabla h = \frac{1}{\rho} \nabla P,
\end{equation}
with $P = \rho^2 \partial e/\partial \rho$ the pressure, one obtains
Euler's equation
\begin{equation}
\label{Euler} 
\partial_0 {\bf v} + \case{1}{2} \nabla {\bf v}^2 + \frac{1}{\rho}
\nabla P = 0
\end{equation}
governing the flow of the fluid.

We next wish to investigate the symmetry content of the theory.   Classical
hydrodynamics has the following invariances:
\begin{itemize}
\item Invariance under space and time translations, $x_\mu \rightarrow x_\mu +
\epsilon_\mu$, with $x_\mu = (t,{\bf x})$, $\mu = 0,1,2,3$ and $\epsilon_\mu$ a
constant vector.
\item Invariance under global translations of the velocity potential, $\theta
\rightarrow \theta + \alpha$, with $\alpha$ a constant.
\item Invariance under Galilei boosts,
\begin{equation} \label{boost} 
t \rightarrow t' = t, \;\; {\bf x} \rightarrow {\bf x}' = {\bf x} -
{\bf u} t; \;\;\;\;
\partial_0 \rightarrow \partial_0' = \partial_0 + {\bf u} \cdot
\nabla, \;\; \nabla \rightarrow \nabla' = \nabla,
\end{equation}    
with ${\bf u}$ a constant velocity.
\end{itemize}
According to Noether's theorem, symmetries imply conservation laws.  More
specifically, the above invariances lead to the conservation laws:
\begin{itemize}
\item $\partial_0 t_{0j} + \partial_i t_{ij} = 0$ and $\partial_0 t_{00} +
\partial_i t_{i0} = 0$, where $t_{\mu \nu}$ $(\mu,\nu = 0,1,2,3)$ is the
energy-momentum tensor, with $t_{0i} = p_i$ the momentum density and $t_{00} =
{\cal H}$ the Hamilton density;
\item $\partial_0 j_0 + \partial_i j_i = 0$, with $j_0$ the mass density
and $j_i$ the mass current;
\item $\partial_0 g_{0j} + \partial_i g_{ij} = 0$, with $g_{0j} = - x_j j_0 +
t \, p_j$ and $g_{ij} = x_i j_i - t \, t_{ij}$ the corresponding charge
densities and currents.  Physically, the conservation $d G_{0i} /dt = 0$ of
the charges $G_{0i} = \int d^3 x \, g_{0i}$ means that the center of mass of
the fluid, ${\bf X} = \int d^3 x \, {\bf x} \rho/ M$, with $M= \int d^3 x \,
\rho$ the total mass, moves with constant velocity,
\begin{equation} 
M \frac{d\bf X}{dt} = \int d^3 x \, {\bf p}.
\end{equation}   
Here, the right-hand side denotes the total momentum of the fluid.
\end{itemize}
From the Lagrange density (\ref{Ltheta}) we obtain as explicit form for the
various charge densities and currents \cite{Kronig}:
\begin{mathletters}
\label{hydro}
\begin{eqnarray}
j_0 &=& -\frac{\partial {\cal L}}{\partial \partial_0 \theta} =
\rho  \\
j_i &=& -\frac{\partial {\cal L}}{\partial \partial_i \theta} =
\rho v_i  \\
t_{i j} &=& {\cal L} \delta_{i j} - \frac{\partial {\cal L}}{\partial
\partial_i \theta} \partial_j \theta =
P \delta_{i j} + \rho v_i v_j \\
p_j (= t_{0 j}) &=& -\frac{\partial {\cal L}}{\partial \partial_0
\theta} \partial_j \theta = \rho v_j  \\
t_{i 0} &=& \frac{\partial {\cal L}}{\partial \partial_i \theta}
\partial_0 \theta = ({\cal H} + P) v_i  \\
{\cal H} (= t_{0 0}) &=& \frac{\partial {\cal L}}{\partial \partial_0
\theta} \partial_0 \theta - {\cal L} =  \frac{\rho}{2} {\bf v}^2  +
\rho e.
\end{eqnarray}
\end{mathletters}
Time derivatives $\partial_0 \theta$ have been eliminated through the
field equation (\ref{rhovar}), so that, for example, 
\begin{equation}  \label{LP}
{\cal L} \rightarrow \rho h - \rho e = \left(\rho \frac{\partial}{\partial
\rho}-1\right) (\rho e) = P.
\end{equation} 
A few remarks are in order.  First, the Hamilton density ${\cal H}$ is the sum
of the kinetic and potential energy density, as required.  Second, the
equivalence of the mass current ${\bf j}$ and the momentum density ${\bf p}$,
which is a hallmark of Galilei invariance, is satisfied by the theory.
Finally, the set of equations (\ref{hydro}) constitutes all the equations of
hydrodynamics.  This brings us to the conclusion that the Lagrange density
(\ref{Ltheta}) encodes all the relevant equations for the description of an
isentropic fluid.

We next turn to the description of sound waves.  We will restrict ourselves to
waves of small amplitude.  These generate only small deviations in the density
$\rho_0$ and pressure $P_0$ of the fluid at rest, so that we can expand the
Lagrange density (\ref{Ltheta}) in powers of $\rho - \rho_0 = \tilde{\rho}$,
with $|\tilde{\rho}| << \rho_0$:
\begin{equation} 
{\cal L} = - (\partial_0 \theta + \case{1}{2}
{\bf v}^2) (\rho_0 + \tilde{\rho}) -  e_0 \rho_0 - h_0 \tilde{\rho}  -
\case{1}{2} h_0' \tilde{\rho}^2  + {\cal O}(\tilde{\rho}^3).
\end{equation}
The derivatives $(')$ with respect to $\rho$ are to be evaluated at
$\rho = \rho_0$.  Since for the system at rest $\theta$ is constant,
it follows from (\ref{rhovar}) that $h_0 = 0$.  If we denote the
thermodynamic derivative $\partial P/ \partial \rho$ by $c^2$, which
has the dimension of a velocity squared, the coefficient of the
quadratic term in $\tilde{\rho}$ can be written as
\begin{equation} 
h_0' =  \frac{1}{\rho_0} P_0' = \frac{c_0^2}{\rho_0}.
\end{equation}
Apart from an irrelevant constant term ($-e_0 \rho_0$) the Lagrange density
becomes to this order
\begin{equation} \label{Lexpand}
{\cal L} = - (\partial_0 \theta + \case{1}{2} {\bf v}^2) (\rho_0 +
\tilde{\rho}) - \frac{c_0^2}{2\rho_0} \tilde{\rho}^2.
\end{equation}
We next eliminate $\tilde{\rho}$ by substituting
\begin{equation}
\label{rhotilde} 
\tilde{\rho} = -\frac{\rho_0}{c_0^2} (\partial_0 \theta + \case{1}{2}
{\bf v}^2),
\end{equation}
which follows from expanding the field equation (\ref{rhovar}).  Physically,
this equation reflects Bernoulli's principle: in regions of rapid flow, the
density $\rho = \rho_0 + \tilde{\rho}$ and therefore the pressure is low.
After eliminating $\tilde{\rho}$, we obtain a Lagrange density governing the
velocity potential $\theta$:
\begin{equation} \label{Leff}
{\cal L}_{\rm eff} = - \rho_0 (\partial_0 \theta + \case{1}{2} {\bf v}^2) +
\frac{\rho_0}{2 c_0^2} (\partial_0 \theta + \case{1}{2} {\bf v}^2)^2.
\end{equation}
This is the standard description of a gapless mode in a non-relativistic
context \cite{GWW,AMJS}.  The effective theory we derived here is apart from
an irrelevant constant identical to a proposal by Takahashi \cite{Takahashi}
which was based on symmetry principles.

The field equation one obtains for the velocity potential $\theta$ from
(\ref{Leff}) is non-linear:
\begin{equation} \label{feq}
\rho_0 (\partial_0^2 \theta + \case{1}{2} \partial_0 {\bf v}^2) - \rho c_0^2
\nabla \cdot {\bf v} + \case{1}{2} \rho_0 ( \partial_0 {\bf v}^2 + {\bf v}
\cdot \nabla {\bf v}^2) =0.
\end{equation} 
The information contained in this equation is nothing more than the
conservation of mass because $\theta$ was initially introduced in (\ref{L}) as
a Lagrange multiplier precisely to enforce this conservation law.  Indeed,
remembering that the combination $\rho_0(\partial_0 \theta + \case{1}{2} {\bf
v}^2)$ denotes $c_0^2$ times $\tilde{\rho}(t, {\bf x}) = \rho(t, {\bf x}) -
\rho_0$, with $\rho_0$ the constant density of the fluid at rest, we see that
(\ref{feq}) reproduces (\ref{mass}) in this approximation.  To simplify
(\ref{feq}), we replace $\rho_0$ in the first and last term by the full
density $\rho$ (which is justified to this order) to arrive at the known
\cite{known}, but unfamiliar field equation
\begin{equation} \label{unfamiliar}
\partial_0^2 \theta - c_0^2 \nabla^2 \theta = - \partial_0 {\bf v}^2 -
\case{1}{2} {\bf v} \cdot \nabla {\bf v}^2
\end{equation} 
governing sound waves in the fluid.  If we ignore the non-linear terms, it
becomes the more familiar wave equation
\begin{equation}
\label{wave}
\partial_0^2 \theta - c_0^2 \nabla^2 \theta = 0,
\end{equation}
implying a gapless linear spectrum, and identifying $c$, which was
introduced via the thermodynamic derivative $\partial P/ \partial \rho
= c^2$, as the sound velocity.  

The combination $\partial_0 \theta + \case{1}{2} (\nabla \theta)^2$ appearing
in the description of a non-relativistic gapless field is dictated by Galilei
invariance.  To obtain the transformation property of the velocity potential
$\theta$ under a Galilei boost (\ref{boost}) we note that since $\nabla
\theta$ is a velocity field, $\nabla \theta(t, {\bf x}) \rightarrow \nabla'
\theta'(t',{\bf x}') = \nabla \theta(t, {\bf x}) - {\bf u}$.  This gives as
transformation rule for $\theta$
\begin{equation}
\label{transfotheta}
\theta(t, {\bf x}) \rightarrow \theta'(t',{\bf x}') = \theta(t, {\bf x}) -
{\bf u} \cdot {\bf x} + f(t),
\end{equation}
with $f(t)$ a yet undetermined function of time.  To determine $f(t)$ we note
that the factor $-\partial_0 \theta$ in the Lagrange density (\ref{Ltheta}) is
the chemical potential per unit mass.  Indeed, using the standard definition
$\mu = \partial {\cal H} / \partial \rho$, we find
\begin{equation}
\mu = \case{1}{2} {\bf v}^2 + h = -\partial_0 \theta,
\end{equation}
where in the second equality we used the field equation
(\ref{rhovar}).  This identification fixes the transformation rule of
$-\partial_0 \theta$:
\begin{equation}
\label{partialtheta}
-\partial_0 \theta(t, {\bf x}) \rightarrow -\partial'_0 \theta'(t', {\bf x}')
= -\partial_0 \theta (t, {\bf x}) - {\bf u} \cdot {\bf v}(t, {\bf x}) +
\case{1}{2} {\bf u}^2
\end{equation}
and in combination with (\ref{transfotheta}), yields for $f(t)$
\begin{equation}
\label{f(t)}
\partial_0 f(t) = \case{1}{2} {\bf u}^2, \;\;\; {\rm or}
\;\;\; f(t) = \case{1}{2} {\bf u}^2 t
\end{equation}
up to an irrelevant constant.  It is easily checked that both the combination
$\partial_0 \theta + \case{1}{2} (\nabla \theta)^2$ appearing in the effective
theory (\ref{Leff}) as well as the field equation (\ref{unfamiliar}) are
invariant under Galilei boosts.  So, contrary to what is sometimes stated in
the literature \cite{Jack}, sound waves are invariant under Galilei boosts.
The wave equation (\ref{wave}) is, of course, not invariant because essential
non-linear terms are omitted.

From the effective theory (\ref{Leff}) one can again calculate the various
Noether charge densities and currents \cite{Takahashi}.  They are, as might be
expected, of the same form as the exact expressions (\ref{hydro}), but now with
the approximations
\begin{equation} 
\label{aprho}
\rho \simeq \rho_0 - \frac{\rho_0}{c_0^2} (\partial_0 \theta +
\case{1}{2} {\bf v}^2)
\end{equation}
as follows from Eq.\ (\ref{rhotilde}),
\begin{equation}
{\cal H} \simeq \frac{\rho}{2} {\bf v}^2 + \frac{c_0^2}{2 \rho_0} (\rho -
\rho_0)^2,
\end{equation}
and
\begin{equation}
P \simeq -\rho_0 ( \partial_0 \theta + \case{1}{2} {\bf v}^2) \simeq c_0^2
(\rho - \rho_0).
\end{equation}
This last equation is consistent with the expression one obtains from
directly expanding the pressure: $P(\rho) = P_0 + \tilde{\rho} P'_0$
since $P_0 = 0$ and $P'_0 = c_0^2$.    

To recapitulate, we have derived an effective theory describing a
massless mode starting from the Lagrange density (\ref{Ltheta}) which
entails the complete hydrodynamics of an  isentropic fluid.  We
now wish to argue that this massless mode is a Goldstone mode
associated with a spontaneously broken symmetry.  A first indication
in favor of such an interpretation follows because the
Hamilton density ${\cal H}$ displays a property typical for a
system with broken symmetry, namely that it is not a function of the
velocity potential itself, but is a function of the gradient of
the field.  The energy is minimal if $\theta$ is uniform throughout
the sample, i.e., there is rigidity \cite{PWA}.  A second indication
comes from the observation that precisely the same effective theory as
(\ref{Leff}) arises in the context of (fermionic and bosonic)
superfluids \cite{AMJS}, and there $\theta$ has been identified as a
Goldstone mode.  

Usually the broken symmetry can be identified by the general property that the
Goldstone mode is translated under the broken symmetry operations.  (There may
be other effects too, but the translation is always there.)  Here, this
general characteristic does not uniquely identify the broken symmetry because
$\theta$ is translated under two symmetry operations.  According to the
transformation rule (\ref{transfotheta}) with $f(t)$ given in (\ref{f(t)}), we
have that under a Galilei boost
\begin{equation}
\label{LeeG}
\delta^{G}_{\bf u} \theta (t, {\bf x}) := \theta' (t, {\bf x}) - \theta (t,
{\bf x}) = - {\bf u} \cdot {\bf x} + t {\bf u} \cdot \nabla \theta (t, {\bf
x}),
\end{equation}
where we took the transformation parameter ${\bf u}$ infinitesimal small so
that quadratic and higher powers in ${\bf u}$ may be ignored.  The first term
at the right-hand side shows that the velocity potential is translated under a
Galilei boost.  The second symmetry under which $\theta$ is translated is
generated by the total mass $M = \int d^3 x \rho$.  To see this we first
compute from (\ref{Ltheta}) the conjugate momentum $\pi_\theta$ of $\theta$
\begin{equation}
\label{pitheta}
\pi_\theta = \frac{\delta {\cal L}_{\rm eff}}{\delta \partial_0 \theta} =
-\rho,
\end{equation} 
and note that $\theta$ and $\rho$ are canonically conjugate \cite{London}:
\begin{equation}
\label{can}
\{\theta(t,{\bf x}), \rho(t,{\bf x}') \} = -\delta ({\bf x} - {\bf x}'),
\end{equation}
where $\{\, ,\}$ denotes the Poisson bracket.  We then use a central result of
classical field theory stating that the charge $Q$ of a continuous symmetry is
the generator of the corresponding transformations of the fields, $\chi (t,
{\bf x}) \rightarrow \chi' (t, {\bf x})$.  More specifically, for an
infinitesimal transformation $\delta_{\alpha}^Q \chi (t, {\bf x}) = \chi' (t,
{\bf x}) - \chi (t, {\bf x})$ one has
\begin{equation}
\delta_{\alpha}^Q \chi (t, {\bf x}) = - \alpha \{\chi (t, {\bf x}), Q\}, 
\end{equation}
with $\alpha$ the transformation parameter.  Equation (\ref{can}) now
immediately shows that under the symmetry generated by the total mass, $\theta$
is indeed translated
\begin{equation}
\delta_{\alpha}^M \theta (t, {\bf x}) = - \alpha \{\theta (t, {\bf x}), M\} =
\alpha.
\end{equation}
The point is that the generators of Galilei boosts $G_{0j} = \int d^3 x (- x_j
j_0 + t p_j)$ also contain the mass density $\rho = j_0$.  It is therefore
impossible to distinguish a broken Galilei invariance from a broken mass
symmetry by considering the algebra alone.

Let us at this point pause and consider the case of a superfluid.  It is well
established that in the normal-to-superfluid phase transition, the mass
symmetry is spontaneously broken.  It is helpful to recall that this comes
about because in a superfluid, which is a quantum system, many particles
Bose-Einstein condense in a single quantum state.  This coherence allows us to
describe the system by a complex field $\psi(t, {\bf x}) = \sqrt{\rho(t, {\bf
x})/m} \exp[i m \varphi(t, {\bf x})/\hbar]$ normalized such that $\psi^* \psi$
yields the particle number density $\rho/m$ of the condensate.  The field
$\varphi$, which will turn out to describe the Goldstone mode of the broken
mass symmetry, is a phase field and therefore compact.  The field $\psi$ is
usually referred to as the condensate wavefunction, even in the modern
literature \cite{AATZ,Stone}.  In our view, this is somewhat misleading.  As
has been stressed by Feynman \cite{Feynman}, $\psi$ is a classical field
describing the coherent behavior of many condensed particles in the same way
as the classical gauge potential of electrodynamics describes the behavior of
many photons in a single state.  For these classical fields there is no
probability interpretation as is required for wavefunctions \cite{Fick}.

It has been argued by Feynman \cite{Feynman} that the $\psi$-field is governed
by a non-relativistic $(\psi^* \psi)^2$-theory defined by the Lagrange density
\begin{equation} \label{Lpsi}
{\cal L}_\psi = i \hbar \psi^* \partial_0 \psi - \frac{\hbar^2}{2m} |\nabla
\psi|^2 - \frac{c_0^2}{2 \rho_0} (m \psi^* \psi - \rho_0)^2.
\end{equation}
The potential energy has its minimum along a circle away from the origin at
$\psi^* \psi = \rho_0$.  This implies a spontaneous breakdown of the mass
symmetry.  The Lagrange density (\ref{Lpsi}) reduces to the one given in
(\ref{Lexpand}) when the term $-\hbar^2 (\nabla \rho)^2/8m^2\rho$ is ignored.
Using the expression for the pressure, cf.\ (\ref{LP})
\begin{equation} 
P = \left[ \rho \left(\frac{\partial }{\partial \rho} -  \partial_i
\frac{\partial }{\partial \partial_i \rho}\right) -1 \right] (\rho e),
\end{equation} 
we find that this term gives the contribution $-\hbar^2 (\nabla^2 \rho)/4m^2$
to the pressure---the so-called quantum pressure.  The reason for calling it
this way is that it is the only place where Planck's constant appears in the
equations.  To the order in which we are working, it is consistent to ignore
this term.  Because the field equation for $\psi$ derived from (\ref{Lpsi})
has the {\em form} of a non-linear Schr\"odinger equation, we will refer to
$\psi$ as a Schr\"odinger field.  We trust however that the reader realizes
that it is a classical field unrelated to a Schr\"odinger wavefunction.

Let us compare the transformation properties of the Schr\"odinger field with
that of the velocity potential $\theta$ of an isentropic fluid.  Under a
Galilei boost, $\psi(t,{\bf x})$ transforms as \cite{Fick}
\begin{equation} \label{Schroeder} 
\psi(t, {\bf x}) \rightarrow \psi'(t', {\bf x}') = \exp[i (-{\bf u} \cdot {\bf
x} + \case{1}{2} {\bf u}^2 t)m/\hbar] \, \psi (t, {\bf x}).
\end{equation} 
With $m \varphi /\hbar$ denoting the phase of the Schr\"odinger field, we see
that $\varphi$ transforms in the same way as does the velocity potential.

Using that the canonical conjugate of the $\psi$-field is $\pi_\psi = i \hbar
\psi^*$, we easily derive the Poisson bracket 
\begin{equation} \label{Poisson}
\{ \psi (t, {\bf x}), M \} = - i (m/\hbar)\psi (t, {\bf x}).
\end{equation}   
This shows that the total mass $M$ generates phase transformations on the
$\psi$-field: $\psi (t, {\bf x}) \rightarrow \psi'(t, {\bf x}) = \exp(i
\alpha m /\hbar) \psi (t, {\bf x})$.  The phase $\varphi$ of the Schr\"odinger
field is consequently translated under the symmetry, just like the velocity
potential $\theta$.  This transformation property identifies $\varphi$ as the
Goldstone mode of the broken mass symmetry.  The Poisson bracket
(\ref{Poisson}) also implies that $\varphi$ and $\rho$ are canonical conjugate
\cite{PWA}, cf.\ (\ref{can})
\begin{equation} 
\{\varphi (t,{\bf x}), \rho(t,{\bf x}') \} = - \delta ({\bf x} - {\bf
x}').
\end{equation} 
A similar relation holds for superconductors.  On quantizing, the Poisson
bracket is replaced by a commutator.  The Heisenberg uncertainty relation that
results for the conjugate pair has recently been demonstrated experimentally
\cite{Delft}.

As we remarked above, a necessary condition for the breaking of the mass
symmetry is the presence of a condensate.  Such an intrinsic quantum
phenomenon, requiring many particles in a single state, has no analog in a
classical setting.  Hence, the mass symmetry cannot be broken in classical
hydrodynamics.  This leaves us with the second possibility, namely that of a
spontaneously broken Galilei invariance.  The breakdown is a result of the
presence of a finite mass density.  This can be inferred \cite{Takahashi} from
considering the transformation of the momentum density ${\bf p}(x)$ under a
Galilei boost:
\begin{equation}
\delta^G_{\bf u} p_i (t, {\bf x}) = -u_j \{p_i(t, {\bf x}), G_{0j}(t) \} =
-u_i \rho (t, {\bf x}) + t {\bf u} \cdot \nabla p_i (t, {\bf x}),
\end{equation}
or with $\delta^G_{\bf u} = u_j \delta^G_j$
\begin{equation}
\delta^G_j p_i (t, {\bf x}) = -\rho (t, {\bf x}) \delta_{j i} + t \partial_j
p_i (t, {\bf x}).
\end{equation}
If the mass density $\rho$ is finite, the left-hand side is non-zero, which is
a symmetry-breaking condition.

There is an essential difference between the breaking of Galilei invariance
and that of mass symmetry.  Although both symmetries are Abelian, the mass
symmetry is a compact U(1) symmetry, whereas the Galilei group is non-compact.
More specifically, the transformation parameter $\alpha$ of the U(1) group has
a finite domain ($0 \leq \alpha < 2 \pi$), while the domain of ${\bf u}$, the
transformation parameter of the Galilei group, is infinite.  This is closely
connected with the impossibility to represent the velocity potential of
classical hydrodynamics as the phase of a complex field.  An immediate
physical manifestation of the difference is that a system with broken U(1)
invariance supports topologically stable vortices, whereas a system with
broken Galilei invariance does not.  This is not to say that vortices are
absent in the latter case, it merely states that their stability is not
guaranteed by topological conservation laws.  Closely connected to this is
that the circulation is not quantized in classical hydrodynamics, which is
known to exist in superfluids.  Yet, the circulation is conserved also in
isentropic fluids.  This is again not for topological, but for dynamical
reasons.  More specifically, the conservation is proven by invoking Euler's
equation (\ref{Euler}) as was first done for an ideal, incompressible fluid by
Helmholtz \cite{Helmholtz} and generalized to a compressible fluid by Thomson
\cite{Thomson}.

The easiest way to observe vortices in a classical fluid is to punch a hole in
the bottom of the vessel containing the fluid.  As the fluid pours out a vortex
is formed in the remaining fluid---a phenomenon daily observed by people
unplugging a sinkhole.  Often, as happens in, for example, superfluid $^4$He,
the core of a vortex is in the normal state so that the symmetry is restored
there.  In the present context this would mean that inside the vortex core, the
fluid mass density $\rho$ is zero.  This is indeed what is observed: the vortex
core consists of air, therefore no fluid is present and $\rho = 0$ there.

In the eye of a tropical cyclone---another example of a vortex, nature
does its best to restore the Galilei symmetry, record low atmospheric
pressures being measured there.  (A complete restoration would imply
the absence of air corresponding to zero pressure.)

It is customary to incorporate vortices in a potential flow via the
introduction of so-called Clebsch potentials \cite{Clebsch}.  We will not
follow this route, but instead use the principle of vortex gauge symmetry
\cite{Kleinert}.  In this approach, one introduces a so-called vortex gauge
field $\theta_\mu^{\rm P} = (\theta_0^{\rm P}, \bbox{\theta}^{\rm P})$ in the
Lagrange density via minimally coupling to the Goldstone field:
\begin{equation} \label{minimal}
\partial_\mu \theta \rightarrow \partial_\mu \theta + \theta_\mu^{\rm P},
\end{equation}
with $\partial_\mu = (\partial_0,-\nabla)$ and 
\begin{equation} \label{vorticity}
\nabla \times \bbox{\theta}^{\rm P} = -2 \bbox{\omega},
\end{equation}
so that $\nabla \times {\bf v} = 2 \bbox{\omega}$ yields (twice) the vorticity
$\bbox{\omega}$ of the vortex.  The object $\theta^{\rm P}_\mu$ describing
the vortex is called vortex gauge field \cite{Kleinert}.  The combination
$\partial_\mu \theta + \theta_\mu^{\rm P}$ is invariant under the local gauge
transformation
\begin{equation} 
\theta(t,{\bf x}) \rightarrow \theta(t,{\bf x}) + \alpha(t,{\bf x}); \;\;\;\;\;
\theta^{\rm P}_\mu \rightarrow \theta^{\rm P}_\mu - \partial_\mu
\alpha(t,{\bf x}),
\end{equation} 
with $\theta^{\rm P}_\mu$ playing the role of a gauge field.  The left-hand
side of (\ref{vorticity}) may be thought of as defining the ``magnetic field''
associated with the vortex gauge field ${\bf B}^{\rm P}= \nabla \times
\bbox{\theta}^{\rm P}$.

For illustrative purposes, let us consider an external, static vortex with
circulation $\Gamma$ located along a line $L$, which may be closed or
infinitely long \cite{Helmholtz}.  Then, $\bbox{\omega} = \case{1}{2}
\Gamma \bbox{\delta } (L)$, where $\bbox{\delta} (L)$ is a delta function on
the line $L$,
\begin{equation} 
\delta_i (L) = \int_L dy_i \, \delta({\bf x} - {\bf y}).
\end{equation} 
This model with a static, external vortex may be thought of as describing the
steady flow in the presence of a vortex pinned to a fixed impurity.  The field
equation for $\theta$ obtained after the substitution (\ref{minimal}) reads
\begin{equation} \label{addeddefect}
\partial_0 \mu + c_0^2 \nabla \cdot {\bf v}
= \partial_0 {\bf v}^2 + \case{1}{2} {\bf v} \cdot \nabla {\bf v}^2 - {\bf
v} \cdot {\bf E}^{\rm P}
\end{equation} 
with $\mu = -(\partial_0 \theta + \theta^{\rm P}_0)$ the chemical potential and
${\bf v} = \nabla \theta - \bbox{\theta}^{\rm P}$ the velocity of the flow in
the presence of the vortex.  The last term gives a coupling of the velocity
field to the ``electric field'' associated with $\theta^{\rm P}_\mu$,
\begin{equation} 
{\bf E}^{\rm P} = -\nabla \theta_0^{\rm P} - \partial_0 \bbox{\theta}^{\rm P}.
\end{equation} 
We note that the field equation (\ref{addeddefect}) is invariant under local
vortex gauge transformations.  Ignoring the higher-order terms and choosing
the gauge $\theta_0^{\rm P} = 0$, we obtain as equation for the flow in the
presence of a static vortex:
\begin{equation}
\nabla \cdot {\bf v} = 0, \;\;\;\; {\rm or} \;\;\;\; \nabla \cdot (\nabla
\theta - \bbox{\theta}^{\rm P}) = 0,
\end{equation}
which is solved by
\begin{equation}
\label{solution}
\theta ({\bf x}) = - \int d^3 y \, G({\bf x} - {\bf y}) \nabla \cdot
\bbox{\theta}^{\rm P}({\bf y}).
\end{equation}
Here, $G({\bf x})$ is the Green function of the Laplace operator
\begin{equation}
G({\bf x}) = \int \frac{d^3 k}{(2 \pi)^3} \frac{ {\rm e}^{i {\bf k}
\cdot {\bf x}}}{{\bf k}^2} = \frac{1}{4 \pi |{\bf x}|}.
\end{equation}
Straightforward manipulations then yield the well-known Biot-Savart
law for the velocity field in the presence of a static vortex
\cite{Helmholtz,Kleinert} 
\begin{equation} \label{vv}
{\bf v}_{\rm v} ({\bf x}) = \frac{\Gamma}{4 \pi} \int_L d {\bf y} \times
\frac{{\bf x} -{\bf y}}{|{\bf x} - {\bf y}|^3},
\end{equation}
where the integration is along the vortex.  This exemplifies the
viability of the vortex gauge principle as an alternative to describe
vortices in a potential flow.  

Let us continue to study the dynamics of vortices---a subject that has recently
received considerable attention in the literature \cite{Sonin}.  Because the
vortex motion is determined by the flow itself, the vortex can no longer be
considered as external.  We shall see that the non-linear part of the field
equation (\ref{addeddefect}) becomes relevant here.

In the absence of external forces, the vortex moves with a constant velocity,
${\bf v}_L$ say.  The flow in the presence of a moving vortex can be obtained
from the static solution (\ref{vv}) by replacing the coordinate ${\bf x}$ with
${\bf x} - {\bf v}_L t$.  This implies that
\begin{equation} \label{vchange}
\partial_0 {\bf v}_{\rm v} = -{\bf v}_L \cdot \nabla {\bf v}_{\rm v} = - \nabla
({\bf v}_L \cdot {\bf v}_{\rm v}).
\end{equation} 
To study sound waves in the presence of a moving vortex, we write the velocity
field as ${\bf v}(t,{\bf x}) = {\bf v}_{\rm v}({\bf x} - {\bf v}_L t) + \nabla
\tilde{\theta}(t,{\bf x})$, with $\tilde{\theta}$ describing small variations
around the moving vortex solution.  Equation (\ref{vchange}) then requires
that we write for the chemical potential in (\ref{addeddefect})
\begin{equation} 
\mu (t, {\bf x}) = {\bf v}_L \cdot {\bf v}_{\rm v}({\bf x} - {\bf v}_L t) -
\partial_0 \tilde{\theta} (t, {\bf x}).
\end{equation}
This leads to the linearized field equation in the frame moving with the
vortex \cite{Sonin}
\begin{equation} \label{AB}
\partial^2_0 \tilde{\theta}(t,{\bf x}) - c_0^2 \nabla^2 \tilde{\theta}(t,{\bf
x}) = - {\bf v}_{\rm v}({\bf x}) \cdot \nabla \partial_0 [2
\tilde{\theta}(t,{\bf x}) - \tilde{\theta}(t,0)],
\end{equation}  
describing sound waves in the presence of a moving vortex.  We have suppressed
the prime on the coordinate ${\bf x}'= {\bf x} - {\bf v}_L t$ of the moving
frame.  In deriving (\ref{AB}) we again used the gauge $\theta^{\rm P}_0=0$,
and replaced in the last term ${\bf v}_L$ with $\nabla
\tilde{\theta}(t,0)$. This is justified because the vortex is carried by the
sound wave and in the moving frame, ${\bf x}=0$ denotes the location of the
vortex.  The first term at the right-hand side in (\ref{AB}) stems from the
non-linear term $\partial_0 {\bf v}^2$ in the general field equation
(\ref{addeddefect}).  Thus the non-linearity of sound waves becomes
detectable.  Equation (\ref{AB}) can be used as a basis to study the
scattering of phonons of a free moving vortex \cite{Sonin}.

So far we have contrasted the breakdown of Galilei invariance (caused by a
finite density) and mass symmetry (caused by a non-zero condensate).  A
superfluid, however, has a finite density as well as a non-zero condensate.
Both symmetries are therefore broken and we expect two different Goldstone
modes to be present.  This is indeed what is observed in superfluid $^4$He.
The system supports besides first sound, which are the usual density waves
associated with the breakdown of Galilei invariance, also second sound, or
entropy waves.  The latter mode depends crucially on the presence of the
condensate and is the Goldstone mode associated with the breakdown of mass
symmetry.  At the transition point, the second sound velocity vanishes,
whereas the first sound velocity remains finite.  This is as expected since
only the condensate vanishes at the superfluid phase transition; the total
mass density remains finite.
\acknowledgements It is a pleasure to thank H. Kleinert and G. Vasconcelos for
useful discussions.
\end{document}